\documentclass[conference]{IEEEtran}
\IEEEoverridecommandlockouts

\ifCLASSINFOpdf
\else
\fi

\usepackage{glossary}
\usepackage{setspace, amsmath, amssymb, url, lscape, subfig, algorithmic, multirow, pslatex, listings, verbatim, alltt, amsfonts, wrapfig, boxedminipage, color, cite}
\usepackage[vlined,linesnumbered,ruled,boxed]{algorithm2e}
\usepackage[dvips]{graphicx}
\usepackage{epsfig}

\usepackage{xcolor}
\usepackage{nicematrix}
\usepackage{array}
\newcommand{\qed}{\nobreak \ifvmode \relax \else
      \ifdim\lastskip<1.5em \hskip-\lastskip
     \hskip1.5em plus0em minus0.5em \fi \nobreak
      \vrule height0.75em width0.5em depth0.25em\fi}

\newcommand{\eg}{{\it e.g., }}
\newcommand{\etal}{{\it et~al., }}
\newcommand{\ie}{{\it i.e., }}
\newcommand{\name}{UMS\xspace}

\newcommand{\comments}[1]{}
\newcommand\hl{\bgroup\markoverwith
  {\textcolor{yellow}{\rule[-.5ex]{2pt}{2.5ex}}}\ULon}
\newcommand{\dind}{Docker-in-Docker\xspace}
\newcommand{\ff}{FastFreeze\xspace}
\newcommand{\init}{\texttt{init}\xspace}
\newcommand{\revised}[1]{{#1}} 

\newcolumntype{C}[1]{>{\centering\arraybackslash}m{#1\textwidth}}

\def\BibTeX{{\rm B\kern-.05em{\sc i\kern-.025em b}\kern-.08em
    T\kern-.1667em\lower.7ex\hbox{E}\kern-.125emX}}
\begin{document}

\title{\name: Live Migration of Containerized Services across Autonomous Computing Systems\\

}


\author{
    \IEEEauthorblockA{
        Thanawat Chanikaphon\\
        HPCC Lab, School of Computing and Informatics\\
        University of Louisiana at Lafayette,
        LA, USA\\
    thanawat.chanikaphon1@louisiana.edu
    }
    \and
    \IEEEauthorblockA{
        Mohsen Amini Salehi\\
        HPCC Lab, Computer Science and Engineering Department\\
        University of North Texas\\
        mohsen.aminisalehi@unt.edu
    }
}


\IEEEaftertitletext{\vspace{-2\baselineskip}}



\maketitle

\IEEEpeerreviewmaketitle
\IEEEaftertitletext{\vspace{-10\baselineskip}}
\begin{abstract}
Containerized services deployed within various computing systems, such as edge and cloud, desire live migration support to enable user mobility, elasticity, and load balancing.
To enable such a ubiquitous and efficient service migration, a live migration solution needs to handle circumstances where users have various authority levels (full control, limited control, or no control) over the underlying computing systems.
Supporting the live migration at these levels serves as the cornerstone of interoperability, and can unlock several use cases across various forms of distributed systems. As such, in this study, we develop a ubiquitous migration solution (called \name) that, for a given containerized service, can automatically identify the feasible migration approach, and then seamlessly perform the migration across autonomous computing systems. \name does not interfere with the way the orchestrator handles containers and can coordinate the migration without the orchestrator involvement. Moreover, \name is orchestrator-agnostic, \ie it can be plugged into any underlying orchestrator platform.
\name is equipped with novel methods that can coordinate and perform the live migration at the orchestrator, container, and service levels. 
Experimental results show that for single-process containers, the service-level approach, and for multi-process containers with small ($<$ 128 MiB) memory footprint, the container-level migration approach lead to the lowest migration overhead and service downtime. To demonstrate the potential of \name in realizing interoperability and multi-cloud scenarios, we examined it to perform live service migration across heterogeneous orchestrators, and between Microsoft Azure and Google Cloud.



\end{abstract}


\section{Introduction}\label{sec:intro}



Applications in smart IoT-based systems, such as those in assistive technologies and autonomous vehicles, often have low-latency constraints to serve their goals.
That is why edge computing has emerged to bypass the network bottleneck and bring the computing to the user (data) proximity, thereby, fulfilling the latency constraints.
The inherent resource shortage and lack of elasticity on the edge, however, has given birth to a new distributed computing paradigm operating based on a continuum of tiers that can include the edge, fog, and cloud systems. To overcome the shortage of edge elasticity, the ability of live service relocation (\ie \emph{service migration}) across the edge-to-cloud continuum is crucial. In addition, enabling service migration can be instrumental in overcoming other longstanding challenges of modern distributed systems, such as user mobility, vendor lock-in, energy efficiency, load balancing, and realizing multi-cloud. 

As an exemplar use case, consider a pair of smartglasses that is used along with the edge-cloud continuum to provide ambient perception for the blind and visually impaired people via real-time services for identification of obstacles and detecting of approaching objects. 
In a hypothetical scenario that is suggestive of the future we hope to create, a blind person enters a coffee shop where people are utilizing the resource-limited on-premise edge server to play an online game. To procure resources for the assistive services of the blind person, the gaming service has to be migrated to the cloud without any significant interruption for the gamers. Migration in the opposite direction can occur when the disabled person leaves the place. In analogy, this is much like a priority seat reserved for disabled people in the public transport systems. Another motivational use case for the live service migration is to avoid vendor lock-in via seamless migration of services across multi-clouds, \ie from one cloud provider to another. 




Provided that modern software engineering methodologies, such as DevOps and CI/CD, predominantly exploit containers \cite{denninnartefficiency} and container orchestrators (\eg Kubernetes)
for service deployments, the key to achieve service migration is to enable the \textit{live migration of the containerized services} across computing systems.
To migrate a containerized service, one may argue that we only need to checkpoint, transfer, and restore the service container.
Indeed, at the high level, this is a valid argument and a container can be transparently checkpointed at the source and transferred to the destination. However, the problem is that, upon container restoration, the destination orchestrator does not recognize and adopt it to offer any management facilities (\eg scaling). The current remedy to this problem (\eg \cite{schrej, tran}) is to make invasive changes to the platform of the underlying computing systems. 
Although the invasive approaches are generally efficient in the sense that they impose a low (lightweight) migration overhead, different computing systems are often controlled autonomously and system administrators do not have the authority to modify both source and destination systems. Moreover, the systems potentially employ distinct orchestrators (\eg Kubernetes and Mesos), whereas, the existing works only perform migration across homogeneous ones that curbs the usability of migration and has vendor lock-in implications. To our knowledge, there is no live service migration solution that can offer the best of both worlds: (i) operating ubiquitously across autonomous systems and heterogeneous orchestrators; and (ii) maintaining the migration efficiency. 


To enable such a ubiquitous and efficient service migration, in this paper, we develop Ubiquitous Migration Solution (\name) that provides migration for different levels of authority the users may have over the underlying computing systems: \emph{full control:} allowing for changes at the platform level of the source and destination systems; \emph{limited control:} allowing changes only to the service image in both systems; and \emph{no control:} that disallows any changes to the underlying systems.  

\name acts as an umbrella solution encompassing the three following migration approaches that correspond to the aforementioned authority levels: (A) \emph{orchestrator-level migration approach} that requires full control over both source and destination systems to be able to make changes in their orchestrator; (B) \emph{service-level migration approach} that demands a limited control only to change the service container image; and (C) \emph{container-level migration approach} that does not demand any control over the underlying systems or services.

Supporting multiple migration approaches raises a challenge within \name to transparently detect the structure of the underlying container and engage the appropriate migration approach. In addition, to support heterogeneous orchestrators, \name has to be able to coordinate the migration across source and destination systems, irrespective of their underlying platforms. Beyond these, \name has to choose the appropriate container(s) for migration. To handle all these complications, we design \name to be a multi-layered such that it can abstract the decision making aspect, from the migration coordination challenges, and from the core migration process.

In summary, this paper makes the following contributions:
\begin{itemize}
\item Developing \name, a framework that enables seamless and lightweight live migration of containerized services across autonomous computing systems with potentially heterogeneous orchestrators\footnote{\small{\name and the experimental data are all available at:\\ \textcolor{blue}{\url{https://github.com/hpcclab/NIMS}}}}.

\item Developing live container migration approaches operating at the orchestrator, container, and service levels. 

\item Demonstrating the feasibility of live migration of containerized services across heterogeneous orchestrators (Kubernetes, Mesos, K3S, and Minishift) and between Microsoft Azure and Google Clouds. We also  analyse the imposed overhead of different migration approaches.
\end{itemize}

The rest of this paper is organized as follows: Section \ref{sec-background} provides a background for the live container migration and its related studies. 
Section \ref{sec-design} presents the design and implementation of \name.
Section \ref{sec:evltn} describes the evaluation and the result.
Finally, Section \ref{sec:conclsn} concludes our work.
\section{Related Works}
\label{sec-background}
Even though the container design principle is often interpreted that containers is ephemeral and migrating persistent storage data \cite{wis23} suffices container migration, many developers and researchers disagree, and several research works have been undertaken to enable the checkpoint/restore of containerized services.
At the orchestrator level, there have been attempts to integrate the checkpoint/restore ability into Kubernetes.
Even though the discussion for this began in 2015 on the Kubernetes GitHub repository 
, there was no tangible outcome until 2020, when Schrettenbrunner presented the proof-of-concept of Kubernetes pod migration in his Ph.D. dissertation \cite{schrej}.
The work requires modifying the source code of the Kubelet and the container runtime interface (CRI) to support the checkpoint/restore operation.
Tran \etal \cite{tran} extended the work and implemented the API server to enable the live migration across two Kubernetes clusters.
Souza \etal \cite{junior2022good} presented MyceDrive, a solution to migrate containers within a Kubernetes cluster based on the service-level migration approach.
To the best of our knowledge, there has been no prior attempt to carry out migration at the orchestrator level based on the container-level migration approach. 

\section{Live Migration of Services Across Autonomous Platforms}
\label{sec-design}

\subsection{Architectural Overview of \name }
\label{subsec:approaches}


\begin{figure}[htbp]
\vspace{-18pt}
\centerline{\includegraphics[width=0.4\textwidth]{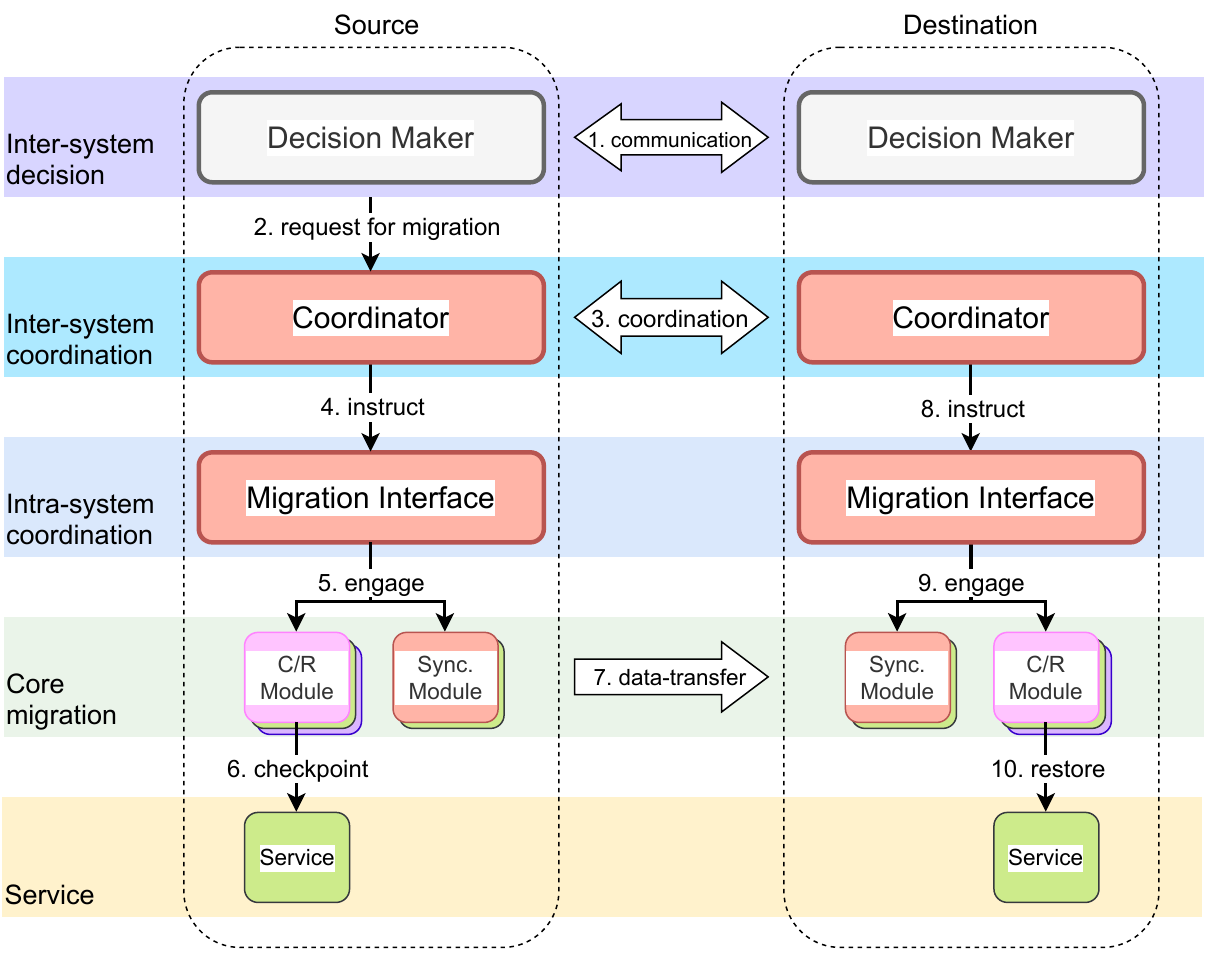}}
\caption{\small{Layered view of the \name architecture. The live migration is performed within the top four layers of the architecture. Red components represent the contributions of this paper.}}
\label{fig:overview}
\vspace{-2pt}
\end{figure}

At the high level, the live container migration consists of five layers, shown in Figure \ref{fig:overview}, namely the \emph{inter-system decision}, \emph{inter-system coordination}, the \emph{intra-system coordination}, the \emph{core migration} and the \emph{service}. \name encompasses the first four layers and the last layer only includes the containerized service, which is incognizant of the migration process.

The \emph{Inter-system decision layer} is responsible for determining the essentials of the live migration process: \emph{What} containerized service(s) is/are the appropriate one(s) for the migration? \emph{Where} should they be migrated to? and \emph{When} is the appropriate time for the migration?
After a decision is made, a request is sent to the \emph{Coordinator} that is accountable for arranging the migration process between the two computing systems. This component determines \emph{``how"} to perform the live migration via transparently identifying the feasible migration approach for the service in question. 
Then, it instructs the \emph{Migration Interface} to engage (call) the modules required to carry out the determined container migration approach. In fact, Migration Interface abstracts the migration coordination from the supported migration approaches. The \emph{Core migration} layer comprises the modules needed to perform the migration procedure, including the \emph{Checkpoint/Restore (C/R) module} and the \emph{Synchronization module} that transfers the checkpoint files to the destination system. 

Although we have implemented the entirety of \name, \textbf{this work concentrates on the live migration mechanism of it} (\ie Inter-and Intra-system coordination layers, and the Core migration layer) to enable seamless and lightweight live service migration across autonomous systems with potentially heterogeneous platforms. At this point, we have placeholders for the \textit{Decision Maker}, such that the migrating container and the destination system are provided as inputs. In the future, we will extend \name to automatically make such decisions.


\subsection{Mechanics of the Live Migration Coordination}
\label{subsec:basic-migration}

\begin{figure*}[htbp]
\centerline{\includegraphics[width=0.8\textwidth]{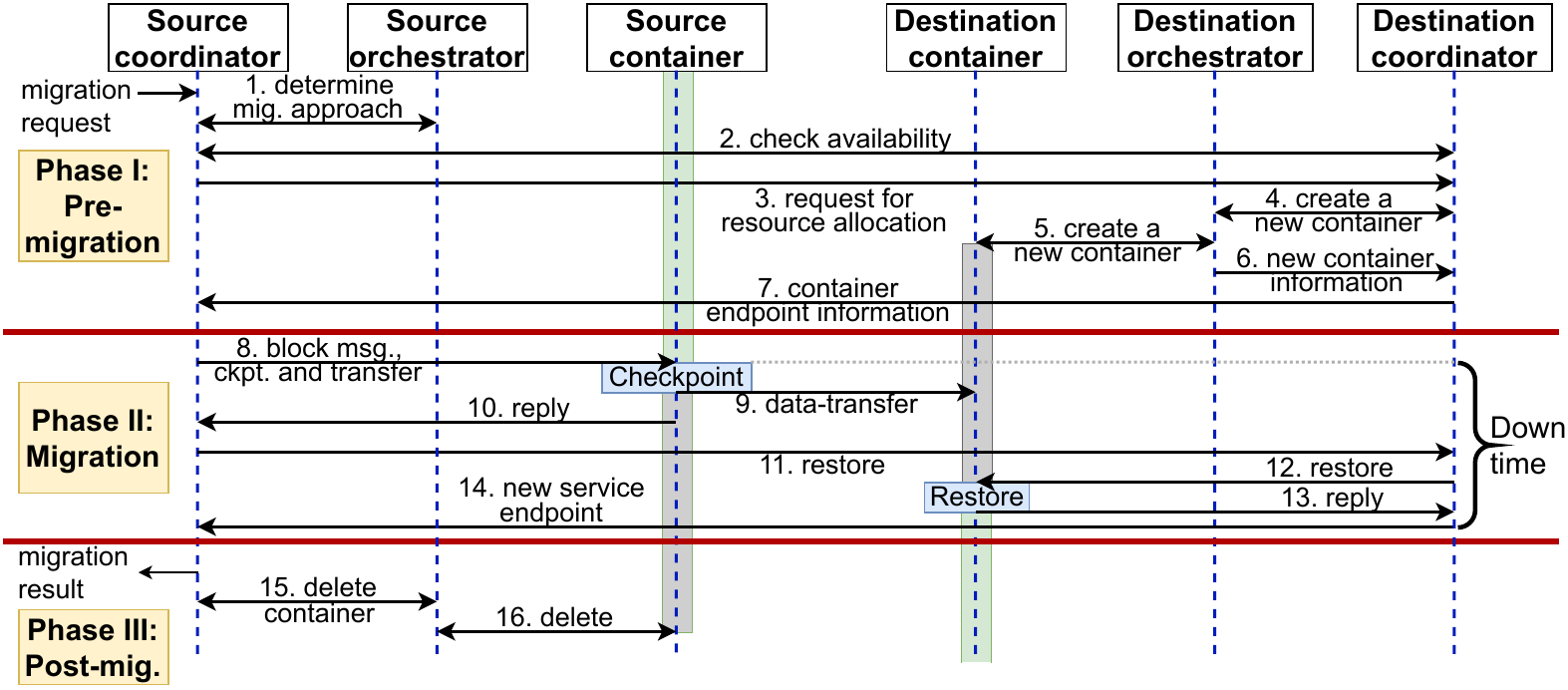}}
\caption{\small{Three phases of live container migration. The bidirectional arrows represent request and reply between entities. The green vertical boxes represent the containerized service in the active state, and the gray ones show them in the inactive state.}}
\label{fig:step}
\vspace{-10pt}
\end{figure*}

Our designed live migration mechanism constitutes three phases that are controlled by the source Coordinator. Figure~\ref{fig:step} elaborates on the sequence of actions in each phase.

\noindent \emph{Pre-migration Phase:}
The migration process begins with the source Coordinator receiving the migration request consisting of two main pieces of information: the containerized service to be migrated; and the destination system to be migrated to. In Step 1, the source Coordinator determines the migration approach in consultation with the orchestrator. In response from the orchestrator, the Coordinator receives the \emph{Specification} of the container in question, including its structure.

In Step 2, a request is sent to verify that the destination Coordinator is available. This step is designed to perform authentication and authorization across systems in the future. Upon confirming the availability, in Step 3, the destination Coordinator is sent the Specification to create a container identical to the one at the source system. Even though, in theory, the destination container creation overhead can be waived via overlapping it with the source container checkpointing step, creating it from early on in the migration process has two benefits: (A) it guarantees the availability and allocation of resources at the destination system for the migration; thus, the migration can be performed safely; and (B) creating the destination container provides the endpoint for the peer-to-peer data-transfer between the source and destination containers in the next phase.
As such, in Steps 4---6, the destination Coordinator creates the new container via its orchestrator and then informs the source Coordinator in Step 7.

\noindent \emph{Migration Phase:}
To avoid inconsistency in the state of the migrating container that can be caused by the arriving messages from the orchestrator, at the beginning of Step 8, the source Coordinator blocks the container from receiving any control messages (\eg deleting). To handle the messages received from the user or other services (\ie data plane), a temporary delegate container, called \emph{Frontman}, is deployed to inform the requester(s) about the temporary service unavailability and asks them to retry. 
Next, the source Coordinator instructs the migrating container to be checkpointed into the storage. It is noteworthy that the reason we use stop-and-copy approach for the migration is that the current container runtimes (\eg Docker) and orchestrators do not support per-copy and post-copy \cite{schrej} approaches.
Although the source container can be terminated after the checkpointing step, to be able to cope with the failures that can occur during the migration, we maintain the source container in an inactive state that is not running and has no dirty pages in the memory until the source Coordinator confirms the safe restoration of the service at the destination. 

In Step 9, the Synchronization module begins to transfer the checkpoint files to the destination container in a peer-to-peer manner. The checkpoint files comprise service memory pages and necessary metadata. The data in persistent storage are transferred concurrently as needed. Finally, upon successful checkpointing, in Step 11, the source Coordinator informs its peer to restore the container from the destination storage.

\noindent \emph{Post-migration Phase:}
After confirming that the service at the destination system started successfully, the source Coordinator informs the migration requester of the new endpoint. Then, in Step 15, the source Coordinator requests the source orchestrator to delete the migrated container. To handle the requests (data plane) received after the migration completion (Step 14), the Frontman container starts redirecting the requests to the new service location. After the DNS is updated with the new endpoint information, the Frontman container is disposed.


\begin{figure}[htbp]
\centerline{\includegraphics[width=0.4\textwidth]{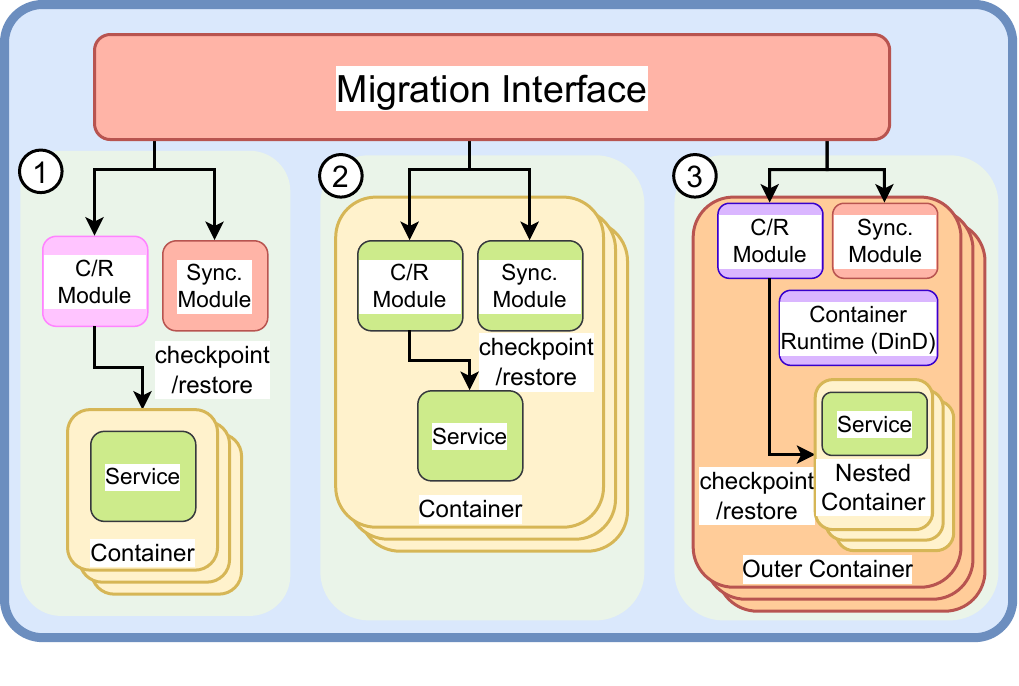}}
\vspace{-8pt}
\caption{\small{Three migration approaches: \textcircled{\raisebox{-.8pt} {1}} orchestrator level, \textcircled{\raisebox{-.8pt} {2}} service level, and \textcircled{\raisebox{-.8pt} {3}} container level.}}
\label{fig:approaches}
\vspace{-10pt}
\end{figure}

\subsection{Establishing Service Migration Approaches Operating at Different Levels}
\label{subsec:architecture}
Once the coordination mechanism is performed, as shown in Figure~\ref{fig:overview}, the Migration Interface is instructed to conduct the core migration via engaging the appropriate approach. For that purpose, Coordinator detects the architecture of the containerized service that itself is dictated by the level of changes we can force to the underlying systems.
Depending on how widely the service is designed to be migrated and the level of authority we have to configure the source and destination systems, as shown in Figure~\ref{fig:approaches}, the following live migration approaches are needed: orchestrator level, service level, and container level.

\paragraph{Orchestrator-level migration approach} To enable this approach, we need to configure the orchestrator of the source and destination systems to be fully compatible. More specifically, the orchestrator should be configured to call the same Checkpoint/Restore and Synchronization modules used by the underlying container runtime. Although this approach is invasive, it enables containers to be efficiently migrated without requiring any modifications.
Tran \etal \cite{tran} developed a live migration solution across two Kubernetes clusters using this approach. They achieved checkpoint/restore via modifying the Kubernetes source code and utilizing Network File System (NFS), a shared storage solution, to transfer the checkpoint files. However, it is not viable across \emph{autonomous} systems, such as those in the edge-cloud and multi-cloud scenarios. 

To enable migration of containerized services without any shared storage, we developed a new synchronization module for \cite{tran} to receive the destination address within the migration request, and transfer the checkpoint files to that address.
To mitigate the migration overhead, we furnished the synchronization module to overlap the container checkpointing and file transfer steps (see Steps 8 and 9 in Figure~\ref{fig:step}), \ie the file transfer step starts without waiting for the checkpointing step to finish.
This is accomplished via monitoring \texttt{write} events in the file system and copying the changed to the checkpoint file.

\paragraph{Service-level migration approach} The orchestrator-level approach demands full control over the underlying orchestrators, however, often our authority is limited and we cannot make changes beyond service images and their deployments. To enable migration under these circumstances, we require a non-invasive migration approach that can function at the service level without any cooperation from the orchestrator. 

To make the service-level migration happen, the Checkpoint/Restore and Synchronization modules must be embedded within the container. As a result, only the \textit{service} memory footprint should be checkpointed and restored within another container at the destination without the need to migrate the entire container.
However, we note that, this approach desires developers to build a migratable container image in both source and destination systems. Moreover, this approach entails developer involvement in the details of the live migration. In this study, we adopted an existing service-level migration solution, known as \ff container \cite{ff}, and extended it to work with the orchestrator. 

\paragraph{Container-level migration approach}
The service-level migration approach is not applicable in circumstances where the user does not have the authority to change the service image. As such, we develop the container-level migration as a non-invasive approach that can perform the live migration in a self-sufficient manner. 

For that purpose, we leverage the ability of container runtimes to perform container checkpoint/restore independent from the orchestrator. However, this ability alone cannot resolve service migration problem, because upon migration, the destination orchestrator does not recognize and evades from managing the migrated service container. To overcome this problem, our solution is to nest the migratable container  within an outer one. On one end, the outer container maintains the binding with the destination orchestrator, and on the other end, it hosts the migrated service as its nested container. 

As shown in part \textcircled{\raisebox{-1pt} {3}} of Figure~\ref{fig:approaches}, the outer container encompasses a container runtime (\eg Docker engine), a Checkpoint/Restore module (\eg CRIU),
and a Synchronization module. This arrangement at the source enables the outer container to migrate its nested one as a nested container of a peer outer container in the destination without any orchestrators' involvement. It is noteworthy that the nested container is just a regular container without any specific adjustments that is managed (\eg in terms of resource usage tracking) by the outer container. To implement the idea of container nesting, we adopt \dind, which is a Docker engine, and deploy it inside the outer container. To synchronize the checkpoint files without any shared storage across systems, we employ the same method explained in the orchestrator-level approach.

\subsection{\name Implementation}
\label{sec-implementation}

We develop the Coordinator and the Migration Interface as web services. Except in the container-level approach, we pack the Migration Interface into the \dind container.
Recall that, in the orchestrator-level approach, the service is a regular container, and in service-level approaches, the service is containerized in \ff-enabled container.
%
%
%
We use Rsync over SSH, 
packed into the Migration Interface, for the data-transfer in the container-level and orchestrator-level approaches. 
\ff has a built-in data transfer tool called \emph{CRIU Image Streamer} 
that can stream the checkpoint files to the destination without buffering them in the local storage. The destination container is configured with MinIO,
an s3 compatible object storage, to receive the checkpoint files. 

\section{Evaluation}
\label{sec:evltn}
Our evaluations encompass three different aspects of the system:
(A) We examine the factors contribute to the latency overhead of each migration approach while varying the sizes of container memory footprints; 
(B) While the other evaluations utilize a benchmarking application with configurable (static) memory footprint, in this part, we inspect the migration performance for a real-world application with a dynamic memory footprint; and 
(C) We study the feasibility of the live migration across heterogeneous orchestrators and multi-clouds.

\label{subsec:evltn-setup}
We created two VMs to simulate cloud-based computing systems, each one in a different physical machine connected by a 1 Gbps link. Each VM includes 8 vCPUs, 16 GiB memory, 50 GiB storage, and a Kubernetes orchestrator.
Lastly, \name is deployed for each orchestrator on each VM.

In most of the experiments, we deployed a popular benchmarking application called \texttt{memhog} \cite{stoyanov2018efficient}. The reason we use \texttt{memhog} is that it can be configured with a static memory footprint, which is a decisive factor in the migration performance. We instruct \texttt{memhog} to allocate a certain amount of memory (in the range of $\approx$0---1,024 MiB), write random data to the allocated memory, and print a counter number at every second. For \ff, we assume that the \ff container image is available at both source and destination systems.

\begin{figure}[htbp]
\vspace{-5pt}
\centerline{\includegraphics[width=0.4\textwidth]{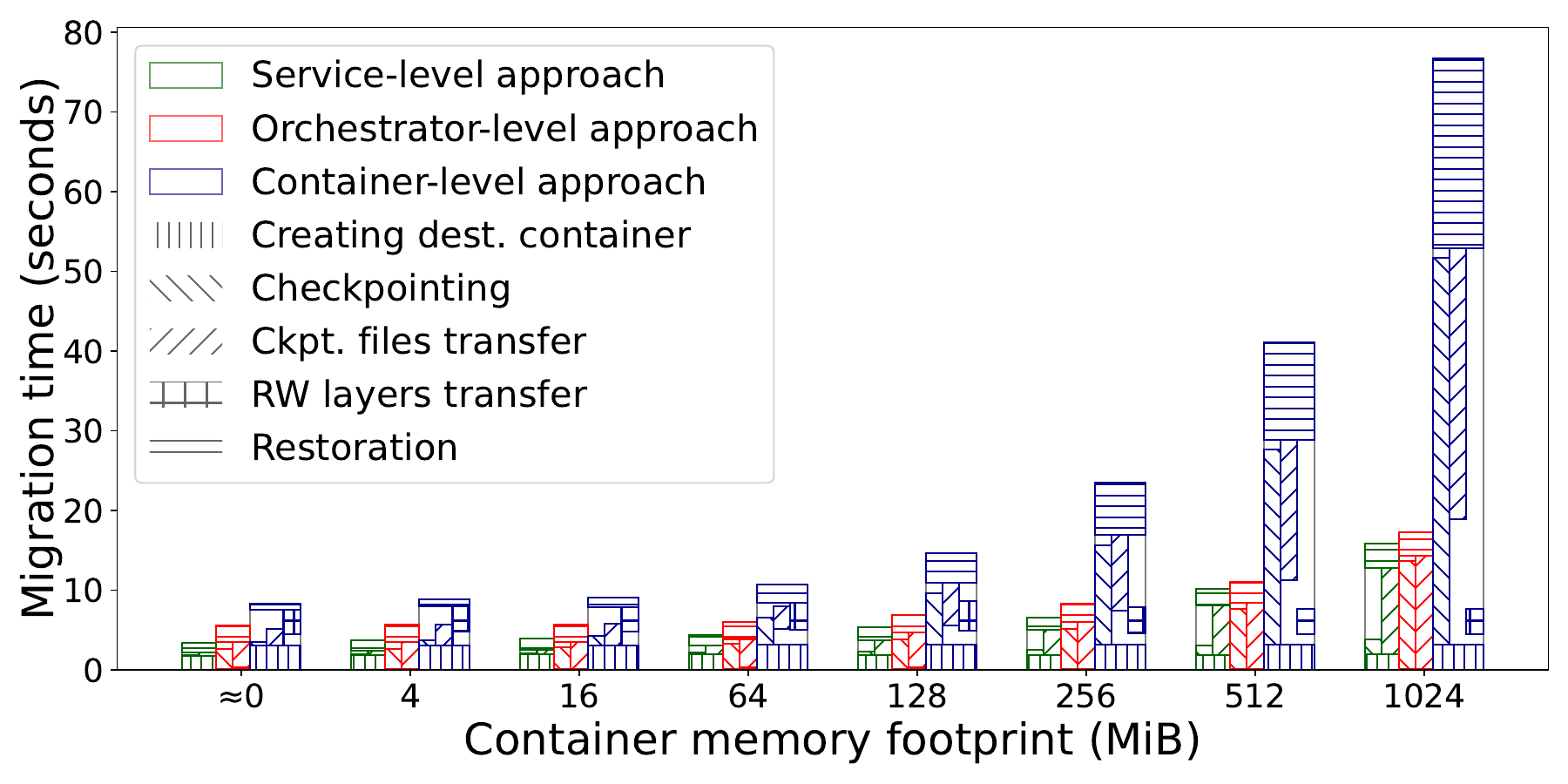}}
\vspace{-5pt}
\caption{\small{Detailed live migration time for different approaches across computing systems. 
\revised{The 90\% confidence interval is negligible.}
}
}
\label{fig:evltn_2}
\vspace{-10pt}
\end{figure}


\subsection{Analyzing the Overhead of Live Container Migration}
\label{subsec:evltn-live-migration}
\vspace{-3pt}
In this experiment, our goal is to measure the overhead of live migration across two homogeneous Kubernetes-based systems upon varying the memory footprint of a single-process container, as shown in Figure~\ref{fig:evltn_2}.
The overhead measurement metric is the migration turnaround time from the migration request at the source until the container runs at the destination. 
We conducted the experiment 30 times and reported the breakdown time of each contributing step to the overall overhead. 

Unsurprisingly, the chart shows that the service-level approach imposes the lowest overhead 
as the overhead of checkpoint, transfer, and restore operations for the entirety of the container is more than doing so only for the service process.
We observe that the orchestrator-level approach outperforms the container-level approach; however, recall that this is an invasive approach that implies changes in the underlying orchestrator, whereas the other two approaches do not.  The overhead difference between the orchestrator-level approach and the container-level approach is $\approx$37\% for small services, \ie containers with less than 128 MiB memory footprint. 
This is an important finding knowing that, in practice, the size of a majority of containerized services is less than 128 MiB.
\revised{In addition, all migration times are considerably high; however, this is common for using cold migration technique\cite{singh2022predictive}.}

\vspace{1mm}
\noindent
\colorbox{blue!10}{
\parbox{0.47\textwidth}{
\underline{\textbf{Takeaway}}: \emph{
For single-process containers, the service-level migration outperforms other approaches. Moreover, the migration time of container-level approach is tolerable for small-size ($<$128 MiB) containers.
}
}}

\begin{figure}[htbp]
\centerline{\includegraphics[width=0.4\textwidth]{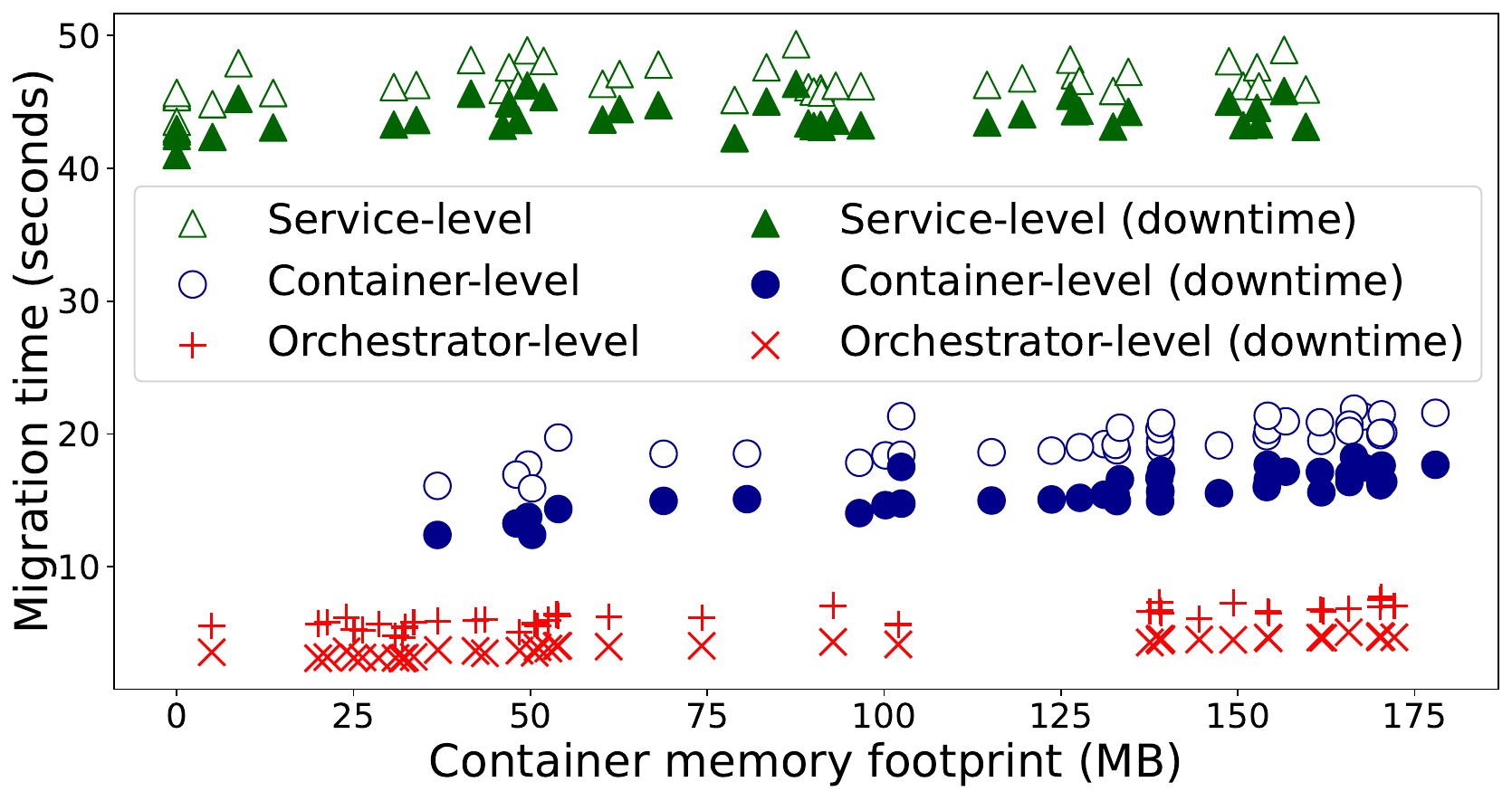}}
\caption{\small{Migration overhead time and service downtime of \texttt{YOLOv3-tiny} for orchestrator-level, container-level, and service-level approaches across Kubernetes orchestrators.}}
\label{fig:evltn_YOLO}
\end{figure}

\begin{figure}[htbp]
\centerline{\includegraphics[width=0.4\textwidth]{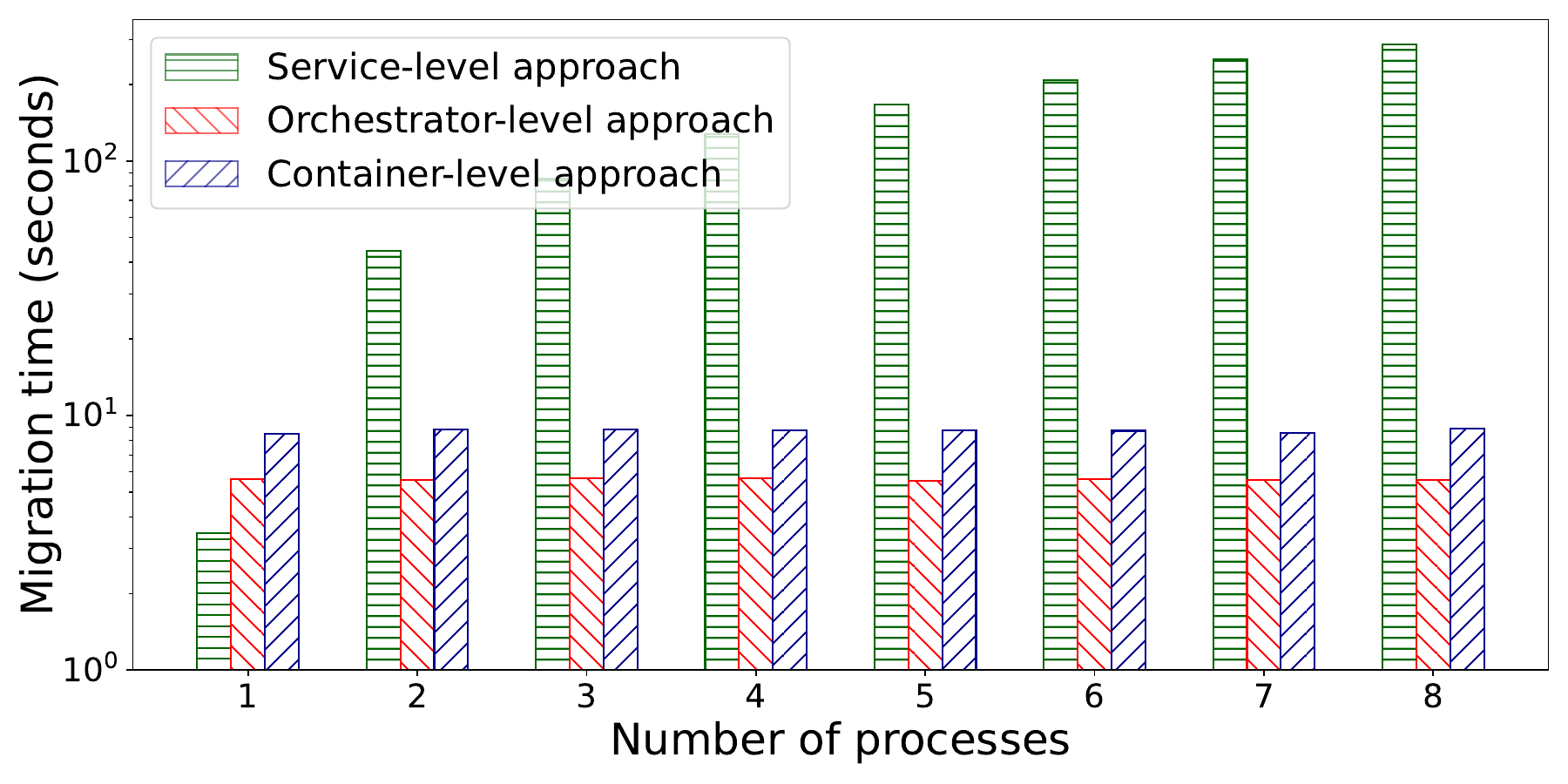}}
\caption{\small{Live container migration time comprises multiple concurrent processes.
\revised{The 90\% confidence interval is $\approx$$\pm$1.0, $\pm$0.1, and $\pm$0.1.}
}}
\label{fig:evltn_3}
\vspace{-10pt}
\end{figure}


\subsection{Impact of Dynamic Memory Footprint on the Migration}
\label{subsec:evltn-YOLO}
In this experiment, we aim to study the migration performance of a service with a dynamic memory footprint. 
We configured \texttt{YOLOv3-tiny} \cite{redmon2018YOLOv3}, a popular object detection application, within a container and fed it with an input image (160 KB) from their repository.
The reason we chose \texttt{YOLOv3-tiny} 
is its predictable (linearly increased) memory footprint behavior upon progress in processing the input image.

To measure the downtime, we conducted the experiment 30 times for each migration approach. At each iteration, we randomly chose a time during the inference process and performed the migration. 
The service downtime was measured by adding a helper process to output a number every second. The downtime is the time interval between the first number printed after the migration and the last number printed before the migration, minus the one-second interval we had by default.

Figure~\ref{fig:evltn_YOLO} demonstrates that the service downtime depends on the migration approach.
We notice that the downtime using the service-level approach is higher than \texttt{memhog} counterparts in Figure~\ref{fig:evltn_2}. 
Our hypothesis is that the reason for the higher downtime is one more process restoration that has to be performed 
for the helper process.
To verify this, we conducted an experiment by configuring \texttt{memhog} to spawn 1---8 child processes with negligible ($\approx$0 MiB) memory footprint, and performed the migration similar to 
Section \ref{subsec:evltn-live-migration}.

Figure~\ref{fig:evltn_3} shows that the service-level approach incurs a significantly higher migration overhead when there is more than one process. 
\revised{
Our analysis shows that this overhead is due to \ff internal mechanics. Specifically, it desires to spawn its child process with a predefined PID, which requires access to the kernel file.
Without privileges, \ff workaround is to keep spawning processes until it reaches the PID it desires, and this process imposes a constant overhead time at the restoration step.
This derives the conclusion that the service complexity (\ie the number of processes running within a container) and privileges are decisive on the downtime of the service-level approach. For such services, even the container-level approach offers a significantly lower migration time.

\vspace{1mm}
\noindent
\colorbox{blue!10}{
\parbox{0.47\textwidth}{
\underline{\textbf{Takeaway}}: \emph{
In practice, container downtime of the service-level approach predominantly depends on the privileges and number of processes running in the container, rather than the container memory footprint at the migration time.
}}}
}

\begin{table}[htbp]
\vspace{-10pt}
\centering
\resizebox{0.5\textwidth}{!}{
\begin{NiceTabular}{C{0.11}|c|c|c|c|c}
 \textbf{Approach} & \Block{}{\textbf{Required} \\ \textbf{changing}} & K8s & Mesos & K3s & Minishift \\
\hline\hline
\Block{}{Orchestrator- \\ level} & Orchestrator & 6.94 & infeasible & infeasible & infeasible \\
\hline
\Block{}{Service- \\ level} & Service & 5.94 & 5.74 & 5.97 & 5.96 \\
\hline
\Block{}{Container- \\ level} & Nothing & 14.71 & 15.03 & 14.63 & 14.85 \\
\end{NiceTabular}
}
\caption{\small{The migration time for a service with 128 MiB memory footprint.
\revised{The 90\% confidence interval is $\approx$$\pm$0.15.}
}}
\label{tab:evltn_3}
\end{table}

\subsection{Live Migration across Heterogeneous Orchestrators}
\label{subsec:evltn-het}
One interesting use case for the live migration of containerized services is to enable heterogeneous orchestrators and multi-cloud scenarios.
This enables enterprises to seamlessly migrate their deployed services to other cloud providers, hence, unlocking the longstanding vendor lock-in problem of the cloud environments, in addition to enabling room for more cost efficiency and service reliability. The service-level and container-level approaches, particularly, fit the public cloud use cases where neither live service migration is supported, nor users are allowed to modify the underlying cloud platforms. As such, the orchestrator-level approach that implies orchestrator-level changes
\revised{and requires compatibility between systems}
is evidently not applicable.
To implement this use case and evaluate it, we developed \name on Mesos,
K3s,
and Minishift, 
and had them migrate the containerized single-process service to Kubernetes. For the sake of better comparison, we also include the case of 
\revised{Section~\ref{subsec:evltn-live-migration}}\footnote{\footnotesize{migrating AKS to GKE demo video is at:} \url{https://youtu.be/SIfIpPWZuls}}.


Table~ \ref{tab:evltn_3} shows the mean migration time for each case after 30 live migration attempts.
Even though the orchestrator-level approach requires changing in the underlying orchestrator, it cannot migrate (shown as \emph{infeasible}) across autonomous systems, \eg multi-cloud, with heterogeneous orchestrators.
Nevertheless, the container-level and the service-level approaches can cover all the cases. The former comes with the benefit of \emph{nothing} to change in the underlying platforms, whereas the latter entails intervention in the service deployment. 

\vspace{1mm}
\noindent
\colorbox{blue!10}{
\parbox{0.47\textwidth}{
\underline{\textbf{Takeaway}:} \emph{Live and seamless migration of containerized services is a viable solution to realize the notion of multi-cloud and resolve the problem of vendor lock-in.}
}}
\section{Conclusion}\label{sec:conclsn}
\vspace{-1pt}
In this research, we developed \name to support seamless and lightweight live migration of containerized services across autonomous systems with potentially heterogeneous orchestrators. \name is equipped with a spectrum of migration approaches (namely, orchestrator-level, service-level, and container-level) that differ in their imposed overhead, and how liberally they can migrate containers across systems.
The results demonstrate that \name can perform low-overhead service migration across any two computing systems. 
\revised{
We concluded that, although the service-level approach is the most lightweight, its performance is downgraded for multi-process containers with insufficient privileges.
We also demonstrated a use case for \name to perform container migration across heterogeneous orchestrators to realize the idea of multi-clouds. 
}
\section*{acknowledgement}
\vspace{-1pt}
This material is supported by National Science Foundation (NSF) under awards\# CNS-2007209 and CNS-2047144.

%
\bibliographystyle{plain} 
\bibliography{references}

\begin{thebibliography}{1}

\bibitem{denninnartefficiency}
Chavit Denninnart, Thanawat Chanikaphon, and Mohsen Amini~Salehi.
\newblock Efficiency in the serverless cloud paradigm: A survey on the reusing
  and approximation aspects.
\newblock {\em Software: Practice and Experience}.

\bibitem{wis23}
Pawissanutt Lertpongrujikorn and Mohsen Amini~Salehi.
\newblock Object as a service (oaas): Enabling object abstraction in serverless
  clouds.
\newblock In {\em Proceedings of the 16th IEEE Cloud Conference}, Jul. 2023.

\bibitem{redmon2018YOLOv3}
Joseph Redmon and Ali Farhadi.
\newblock Yolov3: An incremental improvement.
\newblock {\em arXiv preprint arXiv:1804.02767}, 2018.

\bibitem{schrej}
Jakob Schrettenbrunner.
\newblock {\em Migrating Pods in Kubernetes}.
\newblock PhD thesis, 12 2020.

\bibitem{singh2022predictive}
Gursharan Singh, Parminder Singh, Mustapha Hedabou, Mehedi Masud, and Sultan~S
  Alshamrani.
\newblock A predictive checkpoint technique for iterative phase of container
  migration.
\newblock {\em Sustainability}, 14(11):6538, 2022.

\bibitem{junior2022good}
Paulo Souza~Junior, Daniele Miorandi, and Guillaume Pierre.
\newblock Good shepherds care for their cattle: Seamless pod migration in
  geo-distributed kubernetes.
\newblock In {\em Proceedings of the 6th IEEE International Conference on Fog
  and Edge Computing (ICFEC)}, pages 26--33. IEEE, 2022.

\bibitem{stoyanov2018efficient}
Radostin Stoyanov and Martin~J Kollingbaum.
\newblock {Efficient Live Migration of Linux Containers}.
\newblock In {\em Proceedings of the International Conference on High
  Performance Computing}, pages 184--193, 2018.

\bibitem{tran}
Minh-Ngoc Tran, Xuan~Tuong Vu, and Younghan Kim.
\newblock {Proactive Stateful Fault-Tolerant System for Kubernetes
  Containerized Services}.
\newblock {\em IEEE Access}, 10:102181--102194, 2022.

\bibitem{ff}
Nicolas Viennot.
\newblock {FastFreeze: Unprivileged checkpoint/restore for containerized
  applications}.
\newblock \url{https://lpc.events/event/7/contributions/642/}.
\newblock Online; Accessed on 7 May 2022.

\end{thebibliography}


\end{document}